\documentclass[pra,twocolumn,nofootinbib]{revtex4}
\usepackage{amsfonts,amssymb,amsmath}            
\usepackage[]{graphics,graphicx,epsfig}            
\usepackage{amsthm,bbold}

\newcommand{\ket}[1]{\left | #1 \right\rangle}

\newcommand{\half}{\mbox{$\textstyle \frac{1}{2}$}}

\newcommand{\identity}{\mathbb{1}}
\renewcommand{\epsilon}{\varepsilon}
\begin{document}

\title{The Implications of Ignorance for Quantum Error Correction Thresholds}
\date{\today}

\author{Alastair \surname{Kay}}
\affiliation{Department of Mathematics, Royal Holloway University of London, Egham, Surrey, TW20 0EX, UK}
\begin{abstract}
Quantum error correcting codes have a distance parameter, conveying the minimum number of single spin errors that could cause error correction to fail. However, the success thresholds of finite per-qubit error rate that have been proven for the likes of the Toric code require them to work well beyond this limit. We argue that without the assumption of being below the distance limit, the success of error correction is not only contingent on the noise model, but what the noise model is {\em believed} to be. Any discrepancy must adversely affect the threshold rate, and risks invalidating existing threshold theorems. We prove that for the 2D Toric code, suitable thresholds still exist by utilising a mapping to the 2D random bond Ising model.
\end{abstract}

\maketitle

\section{Introduction}

The inherently analogue nature of quantum superpositions makes the error correction of quantum systems a formidable challenge. In principle, good error correcting codes exist \cite{shor}, and have been implemented experimentally \cite{exp1}. The theoretical extension to the regime of fault-tolerance \cite{ft}, which requires a threshold error rate below which error correction is successful even in the presence of faulty operations, is vital in order to extend existing few-qubit experiments to the realms of functional quantum processors. However, the original route towards fault-tolerance, via a concatenated hierarchy of error correcting codes, introduces such massive overheads that these schemes are impractical for the foreseeable future.

Surface codes, such as the Toric code in 2D \cite{toric}, shift the paradigm of fault-tolerance, disposing of the hierarchical structure. The corresponding reduction in systemic overheads makes them far more promising for experimental implementation. The error correction process only requires classical processing of the results from measurements on small blocks of neighbouring qubits. The resultant error correcting and fault-tolerant thresholds are among the highest known \cite{toric_threshold}, operating in a regime well beyond that predicted by the distance of the code -- for a lattice of $2N^2$ qubits, a logical error can be produced by $\half N$ single qubit errors, while, for large $N$, almost all distributions of $N^2/10$ local errors can be successfully corrected. When operating inside the distance limit of a code, no explicit knowledge of the error mechanism is required. Beyond this limit, two different physical errors may correspond to the same error syndrome. Correcting for the wrong one could lead to a logical error. Evidently, intimate knowledge of the errors is required in order to determine the most likely correction for a given syndrome. However, this information may not be directly available from the syndrome measurements. Previous rigorous derivations of error correcting and fault-tolerant thresholds have assumed perfect knowledge of the noise model (in the sense that a fault-tolerant threshold is derived under the assumption that, for instance, the error rates of two error types are equal). It is therefore important to assess the impact that this ignorance might have on error correcting thresholds. A significant impact is not expected as existing algorithms such as Minimum Weight Perfect Matching \cite{mwpm} function in the biased regime. Nevertheless, our interest here is in proof rather than numerical outcomes from (possibly non-optimal) algorithms running on finite sized systems.

In this paper, we examine the Toric code in 2D, subject to a local noise model. For simplicity of exposition, error correction is assumed to be implemented perfectly, while we might have imperfect knowledge of the error model. We elucidate the influence of making assumptions about the noise model, and prove that the threshold error rates are altered, but not significantly so. Of course, this setting is not physically realistic -- if we can implement the syndrome measurements perfectly, it would seem reasonable that we can also determine the noise mechanism. However, the primary purpose of this paper is simply to convey that one should do this, and, furthermore, should monitor the error model to account for any drift during an experiment. Nevertheless, if that knowledge is imperfect, an error correcting threshold still exists. Moreover, the results presented here can, in principle, be extended to a discussion of fault-tolerance\footnote{The numerical computation of useful bounds becomes a more formidable challenge, although the theory all readily extends.}, at which point we cannot know the error model perfectly.

The main technical tool that we use is the previously established connection between syndrome measurements on a noisy Toric code and the random bond Ising model (RBIM) in 2D \cite{toric_threshold} (or 3D if the measurements are noisy). The phase transition of the RBIM locates the critical threshold of the Toric code. In \cite{toric_threshold}, this connection was established for a noise model parametrised only by $p$, the error rate of a known model. We extend this to a noise model that contains two error rates $\tilde p_X$ and $\tilde p_Z$, and our assumptions about what these values are, $p_X$ and $p_Z$. The critical region of the RBIM can be determined by an ansatz \cite{1,2,takeda,asymmetry} and improved upon by a renormalisation style expansion \cite{corrections}. The values resulting from this ansatz are numerically verified via explicit simulation of a correction algorithm, minimum weight perfect matching.

Other authors \cite{jiannis,lattice_structure} have recently concerned themselves with the idea that two different error types, $X$ and $Z$, could occur at different rates. The standard version of the Toric code in 2D does not tolerate these well, with a threshold of the form $\max(p_X,p_Z)\leq p_C$, and so they have studied how one might alter the lattice geometry in order to better tolerate asymmetries between the parameters $p_X$ and $p_Z$. The model that we choose to study here is a minor transformation of the standard 2D toric code, making it more akin to Wen's model \cite{wen}. This has superior symmetry properties, vastly increasing the range of parameters for which error correction is possible. These thresholds come close to, or even exceed the quantum Hamming bound, which limits the performance of non-degenerate codes by measuring the information content of typical error sequences. While it is known that degenerate codes such as the Toric code can surpass this bound \cite{exceed}, few instances are known.

Before we begin in earnest, let us present an initial statement that justifies our assertions on the existence of good error correcting thresholds. In the presence of $X$ and $Z$ errors, if the two types of error occur independently with the same probabilities $p$, it was shown in \cite{toric_threshold} that error correction is possible if $p<p_C$, some threshold. In that proof, the $X$ and $Z$ errors are treated independently of one another. As such, if the two error rates exhibit some bias, it is clear that error correction is still possible if
$$
\max(p_X,p_Z)<p_C.
$$
So, this immediately shows that a threshold remains if $p_X\neq p_Z$. Moreover, any error correction algorithm that exhibits its own threshold $p<p^A_C$ and operates independently on the two error types must display a similar relationship, $\max(p_X,p_Z)<p^A_C$. This includes, for instance, the Minimum Weight Perfect Matching Algorithm, for which $p^A_C$ is very close to $p_C$, and it achieves this without any knowledge of what the bias is. Hence, perfect knowledge of the underlying noise model is unnecessary. However, that threshold is quite weak: Fig.\ \ref{fig:asymmetrythresholds} compares the small error correcting region (dashed lines) that can be achieved in this way to the quantum Hamming bound (solid black line), which gives a good estimate for how well we might hope to be able to perform. The important issues are how badly the error correcting thresholds are affected by a lack of knowledge of the error model, and to what extent partial knowledge of the error parameters can benefit the threshold.

\subsection{The Toric Code}

The Toric code \cite{toric} is the quintessential example of a surface code. We consider here, as in \cite{kay}, a rotated version akin to the Wen code \cite{wen}. This was also studied in \cite{bombin}. To define it, start from an $N\times N$ square lattice with periodic boundary conditions. Later, it will be convenient for us to term this the {\em primal lattice}. The dual lattice is identical, but shifted by half a lattice vector both horizontally and vertically. On the primal lattice, place a qubit in the middle of each edge. Each vertex $v$ and face $f$ has four neighbouring qubits, two on horizontal edges $E_H$ and two on vertical edges $E_V$. The measurement operators of the code are defined for each vertex and face as
$$
K_v=\prod_{e\in E_H}\!\!Z_e\prod_{e\in E_V}\!\!X_e,\;
K_f=\prod_{e\in E_H}\!\!X_e\prod_{e\in E_V}\!\!Z_e.
$$
All the terms, known as stabilizers, mutually commute and have eigenvalues $\pm1$. The space of Toric code states $\ket{\Psi_{ij}}$ for $i,j\in\{0,1\}$ are defined by the relations $K_v\ket{\Psi_{ij}}=\ket{\Psi_{ij}}$ and $K_f\ket{\Psi_{ij}}=\ket{\Psi_{ij}}$ for all $v,f$. There are $2N^2$ qubits and $2N^2-2$ independent stabilizers, leaving a four-fold degeneracy (the indices $i,j$) that represents two logical qubits. The two logical Pauli $Z^L$ ($X^L$) operators correspond to products of $Z$ ($X$) operations along a single column (row), looping around the entire torus. There are two inequivalent columns (rows), composed of either horizontal or vertical edges. Starting from a logical state $\ket{\Psi_{ij}}$, and applying continuous segments of $X$ and $Z$ operators, it is possible to form closed loops (meaning that all stabilizers return $+1$ expectation). Provided those loops are topologically trivial (i.e.\ they don't form loops around the torus), the state is the same as the original one while non-trivial loops correspond to logical errors. This degeneracy of the code means that if a large set of errors has arisen, it is not necessary to establish exactly which errors occurred in order to correct for them; one only has to form the closed loops which are most likely to be trivial.

We consider an error model of $X$ and $Z$ errors acting independently on each site, with probabilities $\tilde p_X$ and $\tilde p_Z$ respectively: a single qubit state $\rho$ undergoes
$$
\rho\mapsto\mathcal{E}_Z(\mathcal{E}_X(\rho));\quad \mathcal{E}_\sigma(\rho)=(1-\tilde p_\sigma)\rho+\tilde p_\sigma \sigma\rho\sigma.
$$
We will give equivalent results for the extended model
$$
\mathcal{E}(\rho)=(1-\tilde q_X-\tilde q_Z-\tilde q_Y)\rho+\tilde q_XX\rho X+\tilde q_ZZ\rho Z+\tilde q_YY\rho Y
$$
in section \ref{sec:extend}. The restricted error model of $X$ and $Z$ errors only means that there are two independent sets of errors (anyons) that can never interact; those detected by the $\{K_f\}$ and $\{K_v\}$ respectively. By symmetry, it suffices to consider just one of these sets, say $\{K_v\}$. To see this, let us denote by $L$ the lattice on which the toric code is defined (with the qubits in the middles of the edges). Let $L_1$ be a copy of $L$ and $L_2$ be the dual of $L_1$ (i.e.\ in the case of the periodic square lattice, the same lattice but shifted both horizontally and vertically by half a unit). With the edges $q$ (corresponding to a qubit on $L$) of each of the lattices $L_i$ we associate a variable $\tau^i_q\in\pm 1$ in the following way:
\begin{center}
\begin{tabular}{cc|c}
Qubit Type& Error Type & Assignment \\
\hline
$q\in V$ & $X$ & $\tau^2_q=-1$ \\
$q\in V$ & $Z$ & $\tau^1_q=-1$ \\
$q\in H$ & $X$ & $\tau^1_q=-1$ \\
$q\in H$ & $Z$ & $\tau^2_q=-1$
\end{tabular}
\end{center}
Note that $X$ and $Z$ errors affect the two lattices equally\footnote{In the usual (un-rotated) Toric code, all $X$ errors give the values of $\tau^1_q$, and all $Z$ errors give the values of $\tau^2_q$.}.
All other variables $\tau$ are set to 1. Note that this means that $X$ errors are specified by horizontal edges of $L_1$ and $L_2$, while $Z$ errors are specified by vertical edges.

\section{The Random Bond Ising Model}

In \cite{toric_threshold}, a connection was proven between the ability to correct errors arising on the Toric code and the existence of a phase transition in the random bond Ising model (RBIM). This was done for both perfect (the 2D RBIM) and imperfect stabilizer measurement (3D RBIM), assuming that $X$ and $Z$ errors each occur with probability $p$ independently on each lattice site, and assuming that $p$ is known. For simplicity, we will only consider the case of perfect stabilizer measurement, our aim being to relax the assumptions on the knowledge of the error rates, and their equality. To emphasise the difference with the actual error rates ($\tilde p_X$ and $\tilde p_Z$), the assumed error rates are denoted by $p_X$ and $p_Z$ respectively. One should only be able to achieve the optimal recovery specified by \cite{toric_threshold} if the nature of the noise is known exactly.

Whether (or not) the error rates $\tilde p_X$ and $\tilde p_Z$ might be inferred from the syndrome measurements, it is clearly feasible to monitor the error rates to get a good estimate. We do not claim that they should be completely unknown. Rather, our purpose here is twofold: (i) find the error correcting threshold when the rates $p_X=\tilde p_X$ and $p_Z=\tilde p_Z$ and (ii) convey that it is important to be working as close to the conditions $p_X=\tilde p_X$ and $p_Z=\tilde p_Z$ as possible but that, nevertheless, an error correcting threshold still exists, i.e.\ that it is sufficient to have an estimate on the error rates, rather than needing an exact characterisation of the full error model.

For one set of errors, say those affecting the $\{K_v\}$, once the error syndrome has been extracted by measuring the stabilizers, the aim of error correction is to apply a set of operations that reset all the stabilizers to $+1$. Relative to the state that was initially encoded, there are only four inequivalent consequences ($\identity$, $X^L_1$, $Z^L_2$ and $X^L_1Z^L_2$) of the correction. Four corresponding corrections, error strings $E_i$, $i=0\ldots 3$, can be identified and need to be assigned a likelihood of having arisen
$$
p_i=\sum_{C\in S}p(E_i\cup C)
$$
according to the assumptions on the noise model, where the set $S$ corresponds to all trivial loops. If error correction is possible, then the expectation of the probability of getting the right answer over all actual error configurations should tend to 1 in the limit of large system size, $N$, while the other probabilities should vanish. When error correction fails, all the $p_i$ will be similar.

Let $\tau^0$ be a set of variables $\pm 1$ for each qubit, corresponding to whether or not a rotation is applied in the correction $E_0$. Similarly, $\tau^0_C$ is the set due to $E_0\cup C$. The set of all closed loops is conveniently described by introducing variables $\sigma_i\in\{\pm 1\}$ for each vertex of the dual lattice \cite{wuwang,toric_threshold}. A qubit $q$ on an edge of the primal lattice has two neighbouring vertices of the dual lattice, $v^q_1$ and $v^q_2$:
$
\tau^0_{C,q}=\tau^0_q\sigma_{v^q_1}\sigma_{v^q_2},
$
 where it now suffices to sum over the variables $\sigma_i$ without restriction. For probabilities
$$
p_q=\left\{\begin{array}{cc}
p_Z & q\in V\\
p_X & q\in H\\
\end{array}\right.,
$$
we assign the probability of a given error string as
$$
p(\tau_C^0)=\prod_q(1-p_q)^{(1+\tau^0_{C,q})/2}p_q^{(1-\tau^0_{C,q})/2}.
$$
Removing a common factor, we have
$$
\prod_q\left(\frac{1-p_q}{p_q}\right)^{\tau^0_{C,q}/2}.
$$
By defining
$$
\frac{1-p_X}{p_X}=e^{2J_H}\qquad \frac{1-p_Z}{p_Z}=e^{2J_V}
$$
the probability $p_0$ is proportional to
$$
Z_0=\sum_{\vec \sigma}e^{H(\vec\sigma)}
$$
with
\begin{equation}
H(\vec\sigma)=\sum_{q\in H}(\tau^0_q J_H)\sigma_{v^q_1}\sigma_{v^q_2}+\sum_{q\in V}(\tau^0_q J_V)\sigma_{v^q_1}\sigma_{v^q_2}.	\label{eqn:ham}
\end{equation}
This is the Hamiltonian of the $\pm J$ random bond Ising model on a square lattice, where the vector $\tau^0$ arises from the actual errors that occurred, and the coupling strengths constitute our assignment of the likelihood of different configurations. The transition in behaviour of the probabilities between successful correction (in asymptotically all instances of the syndromes) and failure corresponds to a discontinuity of the free energy $F=\ln Z$ of this model, where $Z=\sum_i Z_i$. This is the well studied phase transition in the 2D RBIM. In order to determine the phase transitions, we first establish the duality of the non-random version of the Ising model, and subsequently extend it via the replica method in order to account for the configurational average in the random version.

\begin{figure}[!tb]
\begin{center}
\includegraphics[width=0.2\textwidth]{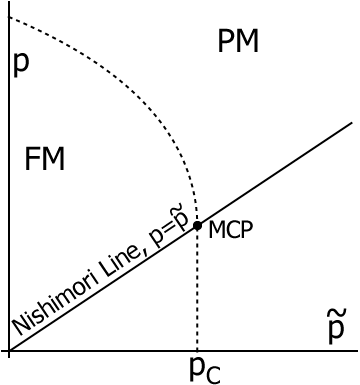}
\end{center}
\vspace{-0.5cm}
\caption{Phase diagram of the 2D RBIM. The dashed line separates the ferromagnetic (FM) and paramagnetic (PM) phases. Where the Nishimori line intersects this is the multicritical point.} \label{fig:phase}
\vspace{-0.5cm}
\end{figure}

\subsection{Duality}

Our analysis starts by considering the non-random bond Ising model, aiming to reproduce the result that the model has a self-dual point \cite{wuwang}. While true for all configurations $\tau^0$, we set $\tau^0_q=1\;\forall q$ for simplicity. Why is duality interesting? It means that as a single parameter $p$ is varied, there's a certain symmetry present such that the model behaves the same also at another point $f(p)$. Assuming the model has exactly one phase transition, the only place that this can occur is the self-dual point $p=f(p)$ (if it occurred at any other point $p$, it would also occur at point $f(p)$). We start by expressing
$$
Z_0=\sum_{\vec{\sigma}}\prod_qu_q(\sigma_{v^q_1}\sigma_{v^q_2}).
$$
This is just writing exactly what we had before but with
$$
u_q(\tau)=\left\{\begin{array}{cc}
e^{J_H\tau}& q\in H \\
e^{J_V\tau}& q\in V
\end{array}\right.
$$
Now, if we define
$$
\sqrt{2}u^*_q(s)=\left\{\begin{array}{cc}
e^{J_H}+(-1)^{s}e^{-J_H}& q\in H \\
e^{J_V}+(-1)^{s}e^{-J_V}& q\in V
\end{array}\right.
$$
for $s\in\{0,1\}$, then we can express
$$
\sqrt{2}u_q(\tau)=\sum_{s_q\in\{0,1\}}u^*_q(s_q)(-1)^{s_q(\epsilon_{v^q_1}+\epsilon_{v^q_2})},
$$
having replaced $\sigma_i$ with $2\epsilon_i-1$ such that $\epsilon_i\in\{0,1\}$. Now consider what happens when we examine the product of $u_q$, expressed as above, for all edges of $L$ leading out of a given vertex $i$, and performing the sum in $Z_0$ over that particular $\epsilon_i$ i.e.
$$
Z_0=\sum_{\vec{\sigma}\setminus \sigma_i}\prod_{\stackrel{r}{v^r_1,v^r_2\neq i}}u_r\sum_{\sigma_i\in\{\pm 1\}}\prod_{q:v^q_1=i}u_q
$$
has that
\begin{eqnarray*}
&\sum_{\sigma_i}\prod_{q:v^q_1=i}u_q(\sigma_i\sigma_{v^q_2})&\\
=&\frac{1}{4}\sum_{\epsilon_i\in\{0,1\}}\prod_{q:v^q_1=i}\sum_{s_q\in\{0,1\}}u^*_q(s_q)(-1)^{s_q(\epsilon_i+\epsilon_{v^q_2})},&
\end{eqnarray*}
and contained within this is
$$
\sum_{\epsilon_i\in\{0,1\}}(-1)^{\epsilon_i\sum_js_j},
$$
which gives a value 2 if
$$
\sum_js_j\text{ mod }2=0,
$$
and 0 otherwise. The sum $j$ is over the vertices such that we include all $q$ with vertices $i$ and $j$. This is entirely equivalent to the product of $s_j$ having to be $+1$ around vertices or, in other words, around closed loops of the dual lattice. Hence,
$$
Z_0=\sideset{}{'}\sum_{s}\prod_qu^*_q
$$
where $^\prime$ denotes the restricted sum only for satisfying assignments around closed loops of the dual lattice\footnote{The factors of 2 conveniently cancel -- for an $n\times m$ lattice, there are $2nm$ edges, and hence a factor $\sqrt{2}^{-2nm}$ appears when replacing the $u$ with $u^*$ for each edge. The compensating factor is the factor of 2 that arises for each vertex (of which there are $nm$) from the sum over $\epsilon_i$.}. This is exactly the same as on the primal lattice, where we could have written
$$
Z_0=\sideset{}{'}\sum_{\tau}\prod_qu_q
$$
with $\tau_q=\sigma_{v^q_1}\sigma_{v^q_2}$, and $^\prime$ indicating a restricted sum only for satisfying assignments around closed loops of the primal lattice. Thus, the model is self-dual when
\begin{equation}
u_q(\tau)=u_q^*((1-\tau)/2)	\label{eq:4eq}
\end{equation}
for all $q$ and all $\tau\in\{\pm 1\}$, remembering that a horizontal edge on the primal lattice corresponds to a vertical edge on the dual lattice. It turns out that the only condition for the self-dual point is
$$
e^{-2J_H}=\tanh(J_V),
$$
satisfying all 4 equations (\ref{eq:4eq}) simultaneously.

\subsection{Replica Method}

The above duality was proven without any randomness present. In order to deal with the randomness of the bonds, we must take a configurational average over the possible values of $\tau^0$, and use it to determine any discontinuity in the free energy, $\ln Z$. The way that we approach calculating this is to consider $n$ parallel copies of the model (all with the same configuration of $\pm 1$ bonds). The partition function of all $n$ parallel copies is just $Z^n$, which is readily calculated for positive integers $n$. If the limit $n\rightarrow 0$ exists, then
$$
\ln Z=\left\langle\lim_{n\rightarrow 0}\frac{Z^n-1}{n}\right\rangle.
$$
For $n$ copies, we can perform the same duality studies as we did above for a single copy. However, the function $u_q(\tau)$ with $\tau\in\{\pm 1\}$ must be replaced with $u_q(\tau)$ with $\tau\in\{\pm1\}^n$, i.e.\ there is a value of $\pm 1$ for the bond $q$ in each copy. The same happens for $u^*$, and self-duality only arises if
$$
u_q(\tau)=u_q^*((1-\tau)/2)
$$
for all $\tau\in\{\pm1\}^n$ simultaneously. This problem reduces slightly because these expressions only depend on the number of the number of $+1$s in the vector $\tau$, so we only need to test equality for $n+1$ cases rather than $2^n$. Let's take a vector $\vec{x}$ of $n+1$ elements, and assign to element $x_p$ the value of $u_q(\tau)$ when $\tau$ contains $p$ $-1$ values.
\begin{eqnarray*}
x_k^H&=&\tilde p_Xe^{(n-2k)J_H}+(1-\tilde p_X)e^{-(n-2k)J_H} \\
x_{k}^{*H}&=&\sqrt{2^{n}}\cosh^{n}(J_H)\tanh^k(-J_H)(1+(1+(-1)^k)\tilde p_X)
\end{eqnarray*}
The $V$ versions are equivalent, with $\tilde p_X\mapsto\tilde p_Z$ and $J_H\mapsto J_V$. Unfortunately, after averaging over the possible random bond assignments $\tau^0_q$, there are no values of $J_H$ and $J_V$ such that $\vec{x}=\vec{x}^*$. Had we been able to, then this is the limit we would have taken as $n\rightarrow\infty$ to find the phase transition. Instead, we follow the approach of \cite{1,2,takeda,asymmetry}, in which it was conjectured that the critical point is approximated by
\begin{equation}
x_0^Hx_0^V=x_0^{*H}x_0^{*V}.	\label{eqn:conjecture}
\end{equation}
Taking the limit $n\rightarrow 0$ yields
\begin{eqnarray}
\tilde p_H\log_2p_H+(1-\tilde p_H)\log_2(1-p_H) && \nonumber\\
+\tilde p_V\log_2p_V+(1-\tilde p_V)\log_2(1-p_V)&=&-1 \label{eqn:zeroorder}
\end{eqnarray}

The ansatz of Eqn.\ (\ref{eqn:conjecture}) was postulated in \cite{1,2,takeda,asymmetry} specifically to work at the multicritical points of the RBIM ($\tilde p_X=p_X$ and $\tilde p_Z=p_Z$), and this was justified by the existence of various symmetries. It has subsequently been numerically tested extensively within this regime, and the asymmetric case of $p_X\neq p_Z$ \cite{asymmetry}. In this case, Eqn.\ (\ref{eqn:zeroorder}) coincides exactly with the quantum Hamming bound. Moreover, the specific instance of $\tilde p_X=\tilde p_Z=p_X=p_Z$ transforms to the Nishimori line of the RBIM, and reveals the critical probability quoted in \cite{toric_threshold} via correspondence to the multicritical point of the RBIM.

\begin{figure}[!tb]
\begin{center}
\includegraphics[width=0.4\textwidth]{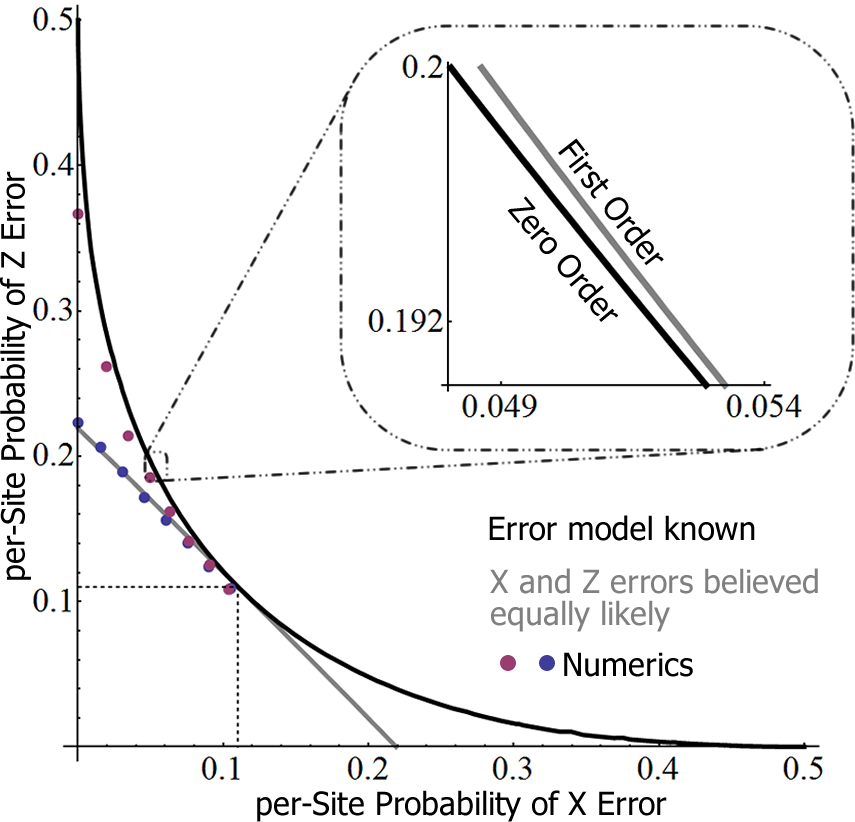}
\end{center}
\vspace{-0.5cm}
\caption{Comparison of error correction threshold when the error model is known ($\tilde p_X=p_X, \tilde p_Z=p_Z$) and when errors are assumed equally likely, $p_X=p_Z$. Plotted for comparison are numerical thresholds from minimum weight perfect matching simulations, see Sec.\ \ref{sec:MWPM}. The enlarged region demonstrates the difference between the zero-order approximation of Eqn.\ (\ref{eqn:zeroorder}), and the first order correction of Eqn.\ (\ref{eq:firstorder}). The dashed region is the error correcting region for the Toric Code in its un-rotated form.} \label{fig:asymmetrythresholds}
\vspace{-0.5cm}
\end{figure}

Practically, we will only ever be able to estimate the parameters $p_X$ and $p_Z$, rather than determine them exactly. Consider the worst possible case, in which we determine the frequency $p$ of stabilizers being $-1$, and assume that $p_X=p_Z$. How detrimental is this to the threshold? Eqn.\ (\ref{eqn:zeroorder}) reduces to an effective homogeneous system with $2\tilde p=\tilde p_X+\tilde p_Z$ and $(1-2p)^2=(1-2\tilde p_X)(1-2\tilde p_Z)$. This point lies on or above the Nishimori line ($\tilde p>p$, see Fig.\ \ref{fig:phase}). While Eqn.\ (\ref{eqn:conjecture}) was not originally proposed to function in this regime, detailed studies \cite{corrections} confirmed that in the homogeneous case, above the Nishimori line, the approximation is a good one. Hence, our analysis remains reliable. Alternatively, having reduced to the homogeneous case, the critical probability cannot be larger than that at the multicritical point, $p_C\approx 0.1092$. Hence, setting $p=\tilde p$ yields
$\tilde p_X+\tilde p_Z< 2p_C$, as compared to the non-transformed version which only successfully corrects if $\max(\tilde p_X,\tilde p_Z)<p_C$. The transformed version has more natural symmetry properties and negates the requirement of recent studies \cite{jiannis,lattice_structure} to adjust the lattice geometry for each different asymmetry between $\tilde p_X$ and $\tilde p_Z$. Fig.\ \ref{fig:asymmetrythresholds} shows that for all parameter values, a finite per-qubit error rate threshold remains and is superior to our original crude estimate (the dashed lines). Provided a sufficiently accurate estimate of the parameters of the error model is made, the error threshold is essentially unaffected.

Although the conjecture of Eqn.\ (\ref{eqn:conjecture}) compares favourably with numerical estimates on the square lattice, there is known to be a discrepancy with some exact renormalisation group calculations on hierarchical lattices \cite{hierarchy1}. To account for this, a renormalisation inspired expansion was introduced in \cite{hierarchy2,corrections} to account for corrections. While most natural for hierarchical lattices \cite{hierarchy2}, it has been extended to square lattices and achieves an even tighter match with previous numerical results using only a first order correction \cite{corrections}. This technique proceeds by replacing the term $e^{J_H}$ in $x^H_0$ (and, similarly, the term $e^{J_H}+e^{-J_H}$ in $x_0^{*H}$) with an equivalent effective weight arising from an averaging effect over several neighbouring spins. The idea is that differing order of correction can be calculated by considering larger and larger neighbourhoods. In the large neighbourhood limit, one must certainly recover the true behaviour of the model. The first order approximation considers only nearest-neighbours, locating the critical point at
\begin{eqnarray}
2&=&\half\sum_{\eta=\pm 1}(1+\eta(1-2p)^4)\log_2(1+\eta(1-2p)^4) \label{eq:firstorder}\\
&&-\sum_{n=0}^1\sum_{m=0}^2\binom{2}{m}a_{nm}(\tilde p_X,\tilde p_Z)\log_2a_{nm}(p_X,p_Z)	\nonumber
\end{eqnarray}
where
$$
a_{nm}(r,s)=\frac{r^ns^m}{(1-r)^{n-2}(1-s)^{m-2}}+\frac{(1-r)^n(1-s)^m}{r^{n-2}s^{m-2}}.
$$
This is a generalised version of Eqn.\ (45) in \cite{corrections}, accounting for an asymmetry between $X$ and $Z$ errors. The threshold values are barely impacted, although Fig.\ \ref{fig:asymmetrythresholds} indicates that they can surpass the zero order approximation. This violation increases at second order, suggesting that this is not a finite sized computational effect, and that this code does indeed have an error correcting threshold (in certain regimes) which exceeds the quantum Hamming bound. This is by no means forbidden (the quantum Hamming bound only applies to non-degenerate error correcting codes, while the Toric code is degenerate), but few examples are known \cite{exceed}.

\section{Minimum Weight Perfect Matching} \label{sec:MWPM}

Given the near-vertical phase boundary of the RBIM below the Nishimori line (Fig.\ \ref{fig:phase}), any error correction strategy which assumes a lower `temperature' has an almost identical critical probability. In particular, the 0 temperature case corresponds to correcting by minimum weight perfect matching \footnote{In this context, temperature is a mathematical parameter of the mapping to the RBIM and has no physical analogue. Zero temperature corresponds to the limit $J\rightarrow\infty$, meaning the only term in $Z_0$ that is worth considering is the $\vec \sigma$ that gives the smallest value.}; an efficient algorithm which is readily implemented. As such, it provides a lower bound on the threshold fidelities for verification of the previous results. For this purpose, the (im)practicalities of its application \cite{poulin} are irrelevant.

The input to a minimum weight perfect matching algorithm is a set of vertices. In the present case, these vertices correspond to the stabilizers which give $-1$ values (i.e.\ the locations of anyons). We must then assign a weight for every pairing of two anyons. We will justify a weighting function momentarily. The algorithm then outputs the way in which the anyons can be paired up (i.e.\ how they might annihilate each other) such that the total weight is minimised. The idea is to make this correction correspond to the most probable set of operations that could have created that distribution of anyons. If a given anyon pair are separated by $l_H$ and $l_V$ in the horizontal and vertical directions, then, as a minimum, they must have been created by $l_H$ $X$ errors and $l_V$ $Z$ errors. Hence, we assign a minimum probability of
$$
\left(\frac{p_X}{1-p_X}\right)^{l_H}\left(\frac{p_Z}{1-p_Z}\right)^{l_V}
$$
to that combination. So, if we take a particular way of pairing up all the anyons, the probability that such a combination arose was the product of all the individual pair-wise probabilities. We want to find the combination that minimises that product, but that's the same as finding the combination that minimises the sum of corresponding logarithms,
$$
l_H\ln\left(\frac{p_X}{1-p_X}\right)+l_V\ln\left(\frac{p_Z}{1-p_Z}\right).
$$
Hence these constitute the weights that we must minimise the total of.

Our simulation, with results depicted in Fig.\ \ref{fig:asymmetrythresholds}, functions by considering an $N\times N$ lattice where $N=100$. We implemented an error model that created $X$ and $Z$ errors on each qubit with probabilities $\tilde p_X$ and $\tilde p_Z$ respectively. Having ascertained the positions of each error (i.e.\ which stabilizers anti-commute with the errors), we assigned weights between a pair of vertices separated by $l_H$ and $l_V$ in the horizontal and vertical directions as
$$
l_H\ln\left(\frac{p_X}{1-p_X}\right)+l_V\ln\left(\frac{p_Z}{1-p_Z}\right).
$$
The Blossom V algorithm \cite{blossom} was then used to perform the minimum weight perfect matching. For a fixed ratio $\tilde p_X/\tilde p_Z$, the fraction of 500 different realisations of an error distribution giving a logical error was computed for varying error rates, enabling determination of the failure probability (the threshold at which a transition in logical error rate from 0 to $50\%$ occurs). Similar numerics, for a perfectly identified error model, are present in \cite{jiannis}.

\section{Generalised Model} \label{sec:extend}

The previous analysis can be repeated for a more general error model of
$$
\mathcal{E}(\rho)=(1-\tilde q_X-\tilde q_Y-\tilde q_Z)\rho+\tilde q_XX\rho X+\tilde q_YY\rho Y+\tilde q_ZZ\rho Z.
$$
Note that we use $q$ to distinguish from the previous $p$. The previous distribution had
$$
q_X=p_X(1-p_Z)\qquad q_Z=p_Z(1-p_X)\qquad q_Y=p_Xp_Z.
$$
With $Y$ errors present, it is not possible to divide the original system into two independent systems, as we did before. Nevertheless, we can still express the probability of successful error correction as being related to the phase transition of the model
$$
Z_0=\sum_{\sigma}e^H
$$
with
$$
H=\sum_{q\in H}\tau^1_qJ_H\sigma_i\sigma_j+\tau^2_qJ_V\sigma_{i'}\sigma_{j'}+\tau^1_q\tau^2_qJ_Y\sigma_i\sigma_j\sigma_{i'}\sigma_{j'}.
$$
where
\begin{eqnarray*}
e^{4J_H}&=&\frac{(1-q_X-q_Y-q_Z)q_Z}{q_Yq_X} \\
e^{4J_V}&=&\frac{(1-q_X-q_Y-q_Z)q_X}{q_Yq_Z} \\
e^{4J_Y}&=&\frac{(1-q_X-q_Y-q_Z)q_Y}{q_Xq_Z}.
\end{eqnarray*}
The $\tau_q^i$ have a sign distribution specified by the error model. This is exactly the model derived in \cite{depolarising}. The duality and replica arguments follow in much the same way. For instance, the duality transformation of a single copy is described by
\begin{eqnarray*}
{\underline u}&=&(e^{J_H+J_V+J_Y},e^{J_H-J_V-J_Y},e^{-J_H+J_V-J_Y},e^{-J_H-J_V+J_Y}) \\
{\underline u}^*&=&\half\left(\begin{array}{cccc}
1 & 1 & 1 & 1\\
1 & -1 & 1 & -1 \\
1 & 1 & -1 & -1 \\
1 & -1 & -1 & 1
\end{array}\right){\underline u},
\end{eqnarray*}
and equality (i.e.\ self-duality) can be generated if, for example, $J_H=J_V$ and
$$
e^{-2J_Y}=\sinh(2J_H).
$$
Finally, we get the equivalent of Eqn.\ (\ref{eqn:zeroorder}), i.e.\ the 0-order approximation, for this generalised model:
\begin{eqnarray*}
(1-\tilde q_X-\tilde q_Y-\tilde q_Z)\log_2(1-\tilde q_X-\tilde q_Y-\tilde q_Z)&&\\
+\sum_{\sigma\in\{X,Y,Z\}}\tilde q_\sigma\log_2\tilde q_\sigma&=&-1.
\end{eqnarray*}

One application is assessing how well minimum weight perfect matching might perform on depolarising noise, as compared to the optimal. If $q_X=q_Y=q_Z=\tilde q_X=\tilde q_Y=\tilde q_Z$, i.e.\ depolarising noise that we have perfectly identified, one has to solve the equation
$$
(1-3q)\log_2(1-3q)+3q\log_2(q)=-1
$$
to find the critical $3q=0.18929$, which replicates the value given in \cite{depolarising}. This is the best that error correction could achieve. What about minimum weight perfect matching? Since it is not capable of taking the correlations introduced by $Y$ into account, it is not expected to be tight with the optimal correction. So, it effectively proceeds by making the assumption that $q_Y=p^2$, $q_X=q_Z=p(1-p)$, $\tilde q_X=\tilde q_Y=\tilde q_Z=\tilde q$. Moreover, we know that the performance of the algorithm is very similar to that of the critical point of the model parametrised in this way. So, it suffices to solve for the critical point again, which is described by
$$
-\half=(1-2q)\log_2(1-p)+2q\log_2(p),
$$
and the largest value of $q$ is given by setting $p=2q$, which reveals that $3q=0.165$. This compares favourably with previous numerical estimates \cite{poulin2}.

\section{Conclusions}

This paper has investigated how any discrepancy between the actual and assumed noise models can influence the error correcting threshold of a code if operating in a regime beyond that specified by the code's distance. The principle is universal, although it has been demonstrated for a specific class of local noise acting on the Toric code in 2D. It is important to note that the rotated version of the Toric code examined here exhibits a significant enhancement in robustness with respect to asymmetries between $X$ and $Z$ error rates, and can even surpass the quantum Hamming bound in some regimes. The concern that the threshold value might be adversely affected, as raised in \cite{kay}, is unfounded; good approximations to the error parameters of a system lead to a negligible change in the threshold error rates. 

We have also detailed how correlations due to $Y$ errors can be incorporated into the analysis \cite{depolarising}. In principle, the present results can be generalised to the situation of noisy measurements \cite{toric_threshold,takeda}, adding a third dimension to model to the RBIM, in order to derive fault tolerant thresholds. It is anticipated that the conclusions of this paper should follow similarly.

\end{document}